\documentclass[%
  10pt,
  nofootinbib,
  superscriptaddress,
  twocolumn,
  amsfonts,
  amssymb,
  amsmath,
  aps,
  pra,
  floatfix
]{revtex4-2}
\usepackage{mathtools} 
\usepackage{mathrsfs}  
\usepackage{amsthm}    
\usepackage{bm}        
\usepackage{braket}    

\usepackage{graphicx}
\usepackage{wrapfig}   
\usepackage{tikz}
\usetikzlibrary{quantikz2}

\usepackage{array}
\usepackage{dcolumn}
\usepackage{multirow}
\usepackage{booktabs}  
\usepackage[inline]{enumitem}  

\usepackage{csquotes}  
\usepackage{comment}   
\usepackage{xcolor}    
\usepackage[normalem]{ulem} 
\usepackage{orcidlink} 

\usepackage{xurl}      
\usepackage{hyperref}
\hypersetup{
  colorlinks=true,
  citecolor=cyan,
  urlcolor=cyan,
  linkcolor=cyan,
  breaklinks=true,
}

\newtheoremstyle{break}
  {} {} {\itshape} {} {\bfseries} {.} {\newline} {\thmname{#1}\thmnumber{ #2}\thmnote{ (\bfseries #3)}}
\theoremstyle{break}

\DeclareMathOperator{\Tr}{Tr}

\let\oldd\d \renewcommand{\d}{\ifmmode\mathrm{d}\else\oldd\fi}
\let\oldi\i \renewcommand{\i}{\ifmmode\mathrm{i}\else\oldi\fi}
\let\Re\relax \DeclareMathOperator{\Re}{Re}
\let\Im\relax \DeclareMathOperator{\Im}{Im}

\newcommand{\expt}[1]{\langle #1 \rangle}
\newcommand{\ketbra}[2]{| {#1} \rangle\langle {#2} |}

\newcommand{\ti}[1]{\textit{#1}}

\newcommand{\ra}{\rightarrow}

\newcommand{\e}{\mathrm{e}}

\newcommand{\eq}[1]{\begin{align} #1 \end{align}}

\makeatletter
\@tfor\Letter:=ABCDEFGHIJKLMNOPQRSTUVWXYZ\do{%
  \expandafter\edef\csname rm\Letter\endcsname{\noexpand\mathrm{\Letter}}
  \expandafter\edef\csname bf\Letter\endcsname{\noexpand\mathbf{\Letter}}
  \expandafter\edef\csname sf\Letter\endcsname{\noexpand\mathsf{\Letter}}
  \expandafter\edef\csname cal\Letter\endcsname{\noexpand\mathcal{\Letter}}
  \expandafter\edef\csname scr\Letter\endcsname{\noexpand\mathscr{\Letter}}
}
\makeatother

\newcounter{one}
{\list{}{\leftmargin=#1\rightmargin=#1}\item[]}%
{\endlist}

\definecolor{lightblue}{rgb}{0.678, 0.847, 0.902}
\definecolor{lightgreen}{rgb}{0.565, 0.933, 0.565}
\definecolor{lightyellow}{rgb}{1.000, 1.000, 0.600}
\definecolor{lightpurple}{rgb}{0.867, 0.627, 0.867}
\definecolor{lightorange}{rgb}{1.000, 0.753, 0.502}
\definecolor{lightpink}{rgb}{1.000, 0.714, 0.757}
\definecolor{lightred}{rgb}{1.000, 0.714, 0.757}
\definecolor{lightcyan}{rgb}{0.878, 1.000, 1.000}
\definecolor{lightcoral}{rgb}{0.941, 0.502, 0.502}
\definecolor{lightsalmon}{rgb}{1.000, 0.627, 0.478}

\newsavebox{\boxA}
\newsavebox{\boxB}
\newsavebox{\boxC}

\begin{document}

\title{Measuring out-of-time-order correlators on a quantum computer\\based on an irreversibility-susceptibility method}

\author{
Haruki Emori
\orcidlink{0009-0007-2264-9192}
}
\email{emori.haruki.i8@elms.hokudai.ac.jp}
\affiliation{
Graduate School of Information Science and Technology, Hokkaido University, Kita 14, Nishi 9, Kita-ku, Sapporo, Hokkaido 060-0814, Japan
}
\affiliation{
RIKEN Center for Interdisciplinary Theoretical and Mathematical Sciences (iTHEMS), RIKEN, 2-1 Hirosawa, Wako, Saitama, 351-0198 Japan
}
\author{
Hiroyasu Tajima
}
\email{hiroyasu.tajima@uec.ac.jp}
\affiliation{
Department of Informatics, Faculty of Information Science and Electrical Engineering, Kyushu University, 744 Motooka, Nishi-ku, Fukuoka 819-0395, Japan
}
\affiliation{
JST, FOREST, 4-1-8, Honcho, Kawaguchi, Saitama 332-0012, Japan
}
\date{\today}

\begin{abstract}
The out-of-time-ordered correlator (OTOC) is a powerful tool for probing quantum information scrambling, a fundamental process by which local information spreads irreversibly throughout a quantum many-body system.
Experimentally measuring the OTOC, however, is notoriously challenging due to the need for time-reversed evolution.
Here, we present an experimental evaluation of the OTOC on a quantum computer, using three distinct protocols to address this challenge: the rewinding time method (RTM), the weak-measurement method (WMM), and the irreversibility-susceptibility method (ISM).
Our experiments investigate the quantum dynamics of an XXZ spin-1/2 chain prepared in a thermal Gibbs state.
As a key contribution, we provide the first experimental demonstration of the ISM, using the trapped-ion quantum computer \texttt{reimei}.
We also conduct a detailed comparative analysis of all three methods, revealing method-dependent behaviors in the measured OTOC.
This work not only validates these protocols as practical tools for exploring quantum chaos on near-term hardware but also offers crucial insights into their respective advantages and limitations, providing a practical framework for future experimental investigations.
\end{abstract}

\maketitle

\section{\label{sec:intro}Introduction}
In recent years, breakthroughs in quantum technology have ushered in a new era of experimental quantum physics.
From ultracold atomic gases and trapped ions to the development of programmable quantum computers, these platforms provide unprecedented opportunities to probe the intricate dynamics of quantum many-body systems far from equilibrium \cite{Altman21,Fauseweh24}.
One of the most compelling frontiers in this field is understanding how information behaves in these complex systems.
While traditional approaches often focus on entanglement and thermalization \cite{Deutsch91,Srednicki94,Rigol08,Polkovnikov11,Kaufman16}, a more refined concept known as quantum information scrambling has emerged as a crucial lens for characterizing non-equilibrium dynamics \cite{Hayden07,Sekino08,Shenker14,Shenker15,Maldacena16JHEP,Hosur16,Roberts17,Swingle17,Lewis-Swan19,Xu24}.

Quantum information scrambling describes the process by which local information becomes irreversibly encoded into non-local degrees of freedom throughout a system.
It is a universal feature of chaotic, strongly interacting quantum systems.
To grasp this concept, one can imagine dropping a single droplet of colored ink into a glass of water.
Initially, the information about the ink droplet's color and position is localized.
As the ink disperses due to diffusion, the information becomes more difficult to recover from any local observation \cite{Bohm80,Zonnios22}.
In a quantum system, unitary time evolution causes a similar effect: an initial local perturbation becomes distributed across the entire system, rendering it unrecoverable by any local measurement.
This is a more subtle process than thermalization, which simply describes how a local subsystem reaches a state of thermal equilibrium with its environment: scrambling however concerns where the \ti{lost} information goes, propagating it to the system's global, non-local degrees of freedom.

The out-of-time-order correlator (OTOC) \cite{Larkin69} has been identified as a uniquely powerful probe for this phenomenon in diverse fields, ranging from high-energy physics to condensed matter physics \cite{Maldacena16JHEP,Swingle17,Lewis-Swan19,Xu24}.
Analogous to the Lyapunov exponent in classical theory of chaos, which measures the sensitivity to initial conditions, the OTOC physically quantifies how a local perturbation spreads or scrambles, throughout a many-body quantum system and mathematically quantifies the growth of a commutator between two initially local operators $W$ and $V$.
At time $t=0$, the commutator $[W(0),V(0)]$ is zero (or small if the operators are spatially separated), as the operators have no shared degrees of freedom.
As time evolves $t=\tau$, however, the time-evolved operator $W(\tau)$ spreads across the system due to interactions.
When its support reaches and overlaps with $V(0)$, the commutator $[W(\tau),V(0)]$ becomes non-zero and its magnitude grows.
This operator spreading is a direct signature of quantum information scrambling.
The rate of this growth provides a clear distinction between different dynamical regimes: chaotic systems typically exhibit an early-time exponential growth, while integrable systems show a much slower, polynomial growth.
This observation makes the OTOC a powerful diagnostic tool for quantum chaos.

Despite its profound theoretical significance, the experimental measurement of the OTOC presents a formidable challenge \cite{Xu24}.
The nature behind the OTOC---\ti{out-of-time-order} of operators---implies a sequence of forward and backward time evolutions, which is notoriously difficult to implement with high fidelity, especially in large, complex, strongly interacting systems \cite{Zangara17,Yan20APR}.
Consequently, much of the research in this area has shifted from theoretical characterization to the development of ingenious experimental protocols that can indirectly access the OTOC, e.g., rewinding time \cite{Swingle16,Swingle18}, interferometer \cite{Zhu16,Yao16}, weak-measurement \cite{Halpern17,Halpern18,Dressel18}, two-point measurement \cite{Campisi17}, R\'{e}nyi entropy \cite{Fan17}, multiple quantum coherence intensity \cite{Wei18}, quantum teleportation \cite{Yoshida19}, randomized measurement \cite{Vermersch19}, thermofield double state \cite{Hurtubise20}, classical shadows \cite{Garcia21}, irreversibility \cite{Emori23} and so forth.

In this paper, we focus on the evaluation of the OTOC using the trapped-ion quantum computer \texttt{reimei}.
We specifically investigate a method based on the principle of irreversibility proposed in the companion Letter \cite{Emori23}, which re-frames the OTOC as the difficulty of recovering an initial state after a specific sequence of operations.
Our main contribution is the first experimental demonstration of this irreversibility-susceptibility method for a non-trivial many-body system, the XXZ spin chain, prepared in a thermal state at finite temperature.
To provide a comprehensive understanding of its performance, we also conduct a comparative analysis of this method against two other prominent techniques: the rewinding time method and the weak-measurement method.
By implementing these distinct methods and analyzing their performance, our work offers crucial insights into their practical applicability on a near-term quantum computer.
This research not only validates quantum computers as powerful platforms for exploring fundamental concepts in quantum chaos but also provides a practical framework for future investigations.

This paper is structured as follows. In Sec.~\ref{sec:otoc}, we provide a detailed mathematical definition of the OTOC and discuss its connection to information scrambling.
In Sec.~\ref{sec:em}, we review three key methods for its evaluation: the rewinding time, weak-measurement, and irreversibility-susceptibility methods.
In Sec.~\ref{sec:ne}, we describe our experimental setup on a quantum computer, including the state preparation algorithm and the Hamiltonian, and present our experimental results.
Finally, in Sec.~\ref{sec:conc}, we conclude with a summary of our findings and a discussion of future prospects.

\section{\label{sec:otoc}Theoretical framework of the OTOC}
The OTOC provides a powerful theoretical framework for quantifying information scrambling and diagnosing quantum chaos in many-body systems \cite{Larkin69,Maldacena16JHEP}.
Its core concept revolves around the dynamics of two initially local operators.

\subsection{Operator spreading and OTOC}
Consider two initially local operators, $W$ and $V$.
In the Heisenberg picture, an operator $W$ evolves in time as $W(\tau):= U^{\dagger}(\tau)W(0)U(\tau)$, where $U(\tau) = \exp(-\i H\tau)$ is the time-evolution operator generated by the Hamiltonian $H$ of the system $\bfS$ under consideration.
While $W(0)$ and $V(0)$ might be simple operators (e.g., a Pauli operator on a single qubit), $W(\tau)$ generally becomes a complex, non-local operator that acts on a larger portion of $\bfS$ as time progresses.
This phenomenon, known as operator spreading, is a key signature of information scrambling.

The OTOC is defined by the thermal expectation value of the squared commutator between $W(\tau)$ and $V(0)$:
\eq{
C_{\beta}(\tau):=-\expt{[W(\tau),V(0)]^{2}}_{\beta}.\label{eq:def_otoc}
}
Here, we abbreviate the Gibbs state $\rho(\beta,H)=\exp(-\beta H)/Z_{\beta}$ and the partition function $Z(\beta,H)=\Tr[\exp(-\beta H)]$ to $\rho_{\beta}$ and $Z_{\beta}$, respectively; $\expt{\bullet}_{\beta}=\Tr(\rho_{\beta}\bullet)$ denotes the thermal average with respect to $\rho_{\beta}$.
This definition is sensitive to the commutator, which is small if the operators are spatially separated, but grows as the time-evolved operator $W(\tau)$ spreads and overlaps with $V(0)$.

A related quantity, the four-point correlator $F_{\beta}(\tau)$, is often studied for its computational convenience:
\eq{
F_{\beta}(\tau):=\expt{W^{\dagger}(\tau)V^{\dagger}(0)W(\tau)V(0)}_{\beta}.\label{eq:def_correlator}
}
If $W$ and $V$ are unitary operators, the squared commutator $C_{\beta}(\tau)$ can be expressed in terms of the real part of $F_{\beta}(\tau)$ as
\eq{
C_{\beta}(\tau)=2\{1-\Re[F_{\beta}(\tau)]\}.\label{eq:otoc-correlator}
}
If, in addition, $W$ and $V$ are also Hermitian, e.g. Pauli operators, this relationship simplifies to $C_{\beta}(\tau)=2[1-F_{\beta}(\tau)]$, where $F_{\beta}(\tau)$ becomes a real quantity.
The squared commutator can also be expressed in terms of the Frobenius norm of the commutator, $C_{\beta}(\tau)=\|\sqrt{\rho_{\beta}}[W(\tau), V(0)]\|^{2}_{2}$.

\subsection{OTOC as a probe of scrambling}
The significance of the OTOC lies in its ability to diagnose operator spreading and chaos.
In a quantum chaotic system, the operator $W(\tau)$ grows in complexity and size. At $t=0$, the commutator $[W(0),V(0)]$ is zero (or small if the operators are spatially separated).
As time evolves $t=\tau$, the support of $W(\tau)$ expands.
When this expanding \ti{operator front} overlaps with the support of $V(0)$, the commutator becomes non-zero, and its magnitude grows \cite{Lieb72}.
This growth signifies the scrambling of quantum information: an initial local perturbation by $V$ becomes entangled with the entire system, and its effect can only be detected by a later, non-local operation $W$ \cite{Qi19,Parker19}.

In systems with a classical chaotic analog, the OTOC often exhibits an early-time exponential growth captured by
\eq{
C_{\beta}(\tau)\sim\e^{2\lambda_{L}\tau},
}
where $\lambda_{L}$ is the quantum Lyapunov exponent \cite{Maldacena16JHEP,Lashkari13,Maldacena16PRD,Zhou19,Xu19}.
This exponential growth marks the initial phase of scrambling.
At long times, the OTOC saturates at a constant value due to the finite size of the system, indicating that the operator has spread across the entire system.

\subsection{Operator expansion and finite temperature}
A more microscopic understanding of the OTOC can be gained by expanding the Heisenberg operator $W(\tau)$ in a complete basis of operators $\{\Gamma_{i}\}$, such as the Pauli strings for a system of qubits:
\eq{
W(\tau)=\sum_{i}\gamma_{i}(\tau)\Gamma_{i}.
}
Here, $|\gamma_{i}(\tau)|^{2}$ can be interpreted as the probability of finding the operator $W(\tau)$ in the state $\Gamma_{i}$.
The thermal average of the OTOC is directly related to these probabilities.
At infinite temperature, the expectation value $\expt{\bullet}_{\beta}$ is replaced by a simple trace, $\Tr(\bullet)/\Tr(\openone)$.
In this case, it can be shown that the average OTOC over all local operators $\Gamma_{i,r}$ at a site $r$ is proportional to the probability that the operator $W(\tau)$ has expanded to include that site as a non-trivial component.
The average squared commutator is then a complementary measure:
\eq{
&\frac{1}{d^{2}-1}\sum_{i:\Gamma_{i,r}\neq\openone_{r}}\frac{1}{\Tr(\openone)}\Tr([W(\tau),\Gamma_{i,r}]^{2})\nonumber\\
&\quad\mbox{}=\frac{2d^{2}}{d^{2}-1}\sum_{i:\Gamma_{i,r}\neq\openone_{r}}|\gamma_{i}(\tau)|^{2},
}
where $d$ is the dimension of the local Hilbert space.
This expression elegantly links the OTOC to the \ti{size} of the Heisenberg operator, which is defined by the number of sites with a non-trivial Pauli operator in its expansion \cite{Xu24}.

While infinite-temperature OTOCs are useful for studying the fundamental physics of scrambling, it is also crucial to consider the finite-temperature case, where the expectation value is taken with respect to a thermal Gibbs state $\rho_{\beta}$.
The physics in this regime is more complex due to correlations present in the thermal state.
For instance, there are multiple \ti{regularized} versions of the measure for the OTOC, e.g.,
\eq{
\tilde{F}_{\beta}(\tau):=\Tr[\rho^{\kappa_{1}}W(\tau)\rho^{\kappa_{2}}V(0)\rho^{\kappa_{3}}W(\tau)\rho^{\kappa_{4}}V(0)],
}
where $\sum_{i}\kappa_{i}=1$.
These different regularizations correspond to displacing the operators in imaginary time and can lead to different physical results, such as a dependence of the butterfly velocity on the choice of the regularization \cite{Hosur16,Qi19,Parker19,Blake18,Liao18,Chan19,Murthy19,Romero-Bermúdez19,Sahu20,Pappalardi22,Nezami23}.
Nevertheless, the core principle of the OTOC as a probe of operator spreading remains intact, providing a powerful tool for diagnosing information scrambling across various physical regimes.

\def\myvdots{\ \vdots\ }
\savebox{\boxA}{
\resizebox{0.9\linewidth}{!}{
\begin{quantikz}[wire types = {q,q,q,n,q,q}, thin lines,] 
\lstick{$\ketbra{+}{+}_{\bfC}$} & \ctrl{1} & \qw & \qw & \qw & \ctrl[open]{1} & \meterD{X/Y} \\
\lstick[5]{$\rho_{\mathbf{S}}$} & \gate{V(0)} & \gate[5]{U(\tau)} & \qw & \gate[5]{U^{\dagger}(\tau)} & \gate{V(0)} & \\
&&&&&&\\
\myvdots & & & \myvdots & & & \myvdots\\
&&&&&&\\
\qw & \qw & \qw & \gate{W(0)} & \qw & \qw & \qw
\end{quantikz}
}}

\savebox{\boxB}{
\resizebox{1.4\linewidth}{!}{
\begin{quantikz}[wire types = {q,q,q,q,n,q,q,q,q}, thin lines] 
\lstick{$\ketbra{+}{+}_{\bfP_{v'}}$} & \qw & \qw & \qw & \qw & \gate[3]{S_{V}(\phi_{v'})} & \meterD{Z} \\
\lstick{$\ketbra{+}{+}_{\bfP_{v}}$} & \gate[2]{S_{V}(\phi_{v})} & \meterD{Z} \\
\lstick[5]{$\rho_{\mathbf{S}}$} & \qw & \gate[5]{U(\tau)} & \qw & \gate[5]{U^{\dagger}(\tau)} & \qw & \gate[5]{U(\tau)} & \qw & \qw & \qw \\
&&&&&&&&&\\
\myvdots & & & \myvdots & & \myvdots & & & & \myvdots \\
&&&&&&&&&\\
\qw & \qw & \qw & \gate[2]{S_{W}(\phi_{w})} & \qw & \qw & \qw & \qw & \gate[3]{S_{W}(\phi_{w'})} & \qw \\
\lstick{$\ketbra{+}{+}_{\bfP_{w}}$} & \qw & \qw & \qw & \meterD{Z} \\
\lstick{$\ketbra{+}{+}_{\bfP_{w'}}$} & \qw & \qw & \qw & \qw & \qw & \qw & \qw & \qw & \meterD{Z}
\end{quantikz}
}}

\savebox{\boxC}{
\resizebox{0.96\linewidth}{!}{
\begin{quantikz}[wire types = {q,q,q,n,q,q}, thin lines]
\lstick{$\ketbra{\pm}{\pm}_{\bfQ}$} & \gate[2]{U_{V}(\theta)} & & & & \gate[2]{U^{\dagger}_{V}(\theta)} & \meterD{X} \\
\lstick[5]{$\rho_{\mathbf{S}}$} & & \gate[5]{U(\tau)} & & \gate[5]{U^{\dagger}(\tau)} & & \\
&&&&&&\\
\myvdots & & & \myvdots & & & \myvdots \\
&&&&&&\\
& & & \gate{W(0)} & & &
\end{quantikz}
}}

\section{\label{sec:em}Evaluation methods}
While the experimental evaluation of the OTOC is a non-trivial task, each method presents unique theoretically interesting advantage and experimental challenges.
We now explain these experimental evaluation methods of the OTOC, specifically focussing on the rewinding time method (RTM) \cite{Swingle16,Swingle18}, weak-measurement method (WMM) \cite{Halpern17,Halpern18,Dressel18}, and irreversibility-susceptibility method (ISM) \cite{Emori23} for the comparison in our numerical experiments.

\subsection{\label{subsec:rtm}Rewinding time method}
\begin{figure}[tb]
\centering
\usebox\boxA
\caption{
Schematic diagram of the rewinding time method (RTM).
The RTM is a procedure to evaluate the four-point correlator $F_{\beta}(\tau)$.
The process begins with a controlled operation $\ketbra{0}{0}_{\bfC}\otimes\openone_{\bfS}+\ketbra{1}{1}_{\bfC}\otimes V(0)_{\bfS}$.
This is followed by a forward time evolution $\openone_{\bfC}\otimes U(\tau)_{\bfS}$, an operation $\openone_{\bfC}\otimes W_{\bfS}(0)$, and a backward time evolution $\openone_{\bfC}\otimes U^{\dagger}(\tau)_{\bfS}$.
The sequence concludes with a final controlled operation $\ketbra{0}{0}_{\bfC}\otimes V(0)_{\bfS}+\ketbra{1}{1}_{\bfC}\otimes\openone_{\bfS}$.
Finally, the control system $\bfC$ is measured in both the $\sigma^{x}$ and $\sigma^{y}$ bases to calculate $F_{\beta}(\tau)$.
}\label{fig:rtm}
\end{figure}

The RTM \cite{Swingle16,Swingle18} provides a direct way to measure the OTOC by relating it to the outcome of a quantum interference experiment \cite{Zangara17,Yan20APR}.
The central idea of this method is to measure the overlap between two quantum states, which are prepared by applying local operators and time evolution in different orders.

Let the target system $\bfS$ be initialized in a pure state $\ket{\psi}$ for simplicity.
We are interested in the overlap of the states
\eq{
\ket{\psi_{1}}&=V(0)W(\tau)\ket{\psi},\\
\ket{\psi_{2}}&=W(\tau)V(0)\ket{\psi}.
}
The OTOC is related to this overlap, and an interferometric setup allows us to precisely quantify the relationship.

To measure $F_{\beta}(\tau)$, we employ a practical quantum circuit that uses an ancillary qubit $\bfC$ as a control system.
\begin{enumerate}
\item Prepare $\bfC$ in the eigenstate $\ket{+}_{\bfC} = (\ket{0}_{\bfC}+\ket{1}_{\bfC})/\sqrt{2}$ of the Pauli-$x$ operator $\sigma^{x}$, which corresponds to the eigenvalue $+1$, and $\bfS$ in the desired state $\ket{\psi}_{\bfS}$.
The initial state of composite system $\bfC+\bfS$ is $\ket{\Psi_{\text{in}}}=\ket{+}_{\bfC}\otimes\ket{\psi}_{\bfS}$.
\item Apply the controlled-operator $\ketbra{0}{0}_{\bfC}\otimes\openone_{\bfS}+\ketbra{1}{1}_{\bfC}\otimes V(0)_{\bfS}$.
$V(0)$ is a local operator, meaning it acts only on a specific part of $\bfS$.
\item Apply the forward time evolution $\openone_{\bfC}\otimes U(\tau)_{\bfS}$.
\item Apply the operator $\openone_{\bfC}\otimes W_{\bfS}$.
\item Apply the backward time evolution $\openone_{\bfC}\otimes U^{\dagger}(\tau)_{\bfS}$.
\item Apply the controlled-operator $\ketbra{0}{0}_{\bfC}\otimes V(0)_{\bfS}+\ketbra{1}{1}_{\bfC}\otimes\openone_{\bfS}$.
\end{enumerate}
After these operations, the state of $\bfC+\bfS$ becomes an entangled state
\begin{equation}
\ket{\Psi_{\text{out}}}=\frac{1}{\sqrt{2}}\left(\ket{0}_{\bfC}\ket{\psi_{1}}_{\bfS}+\ket{1}_{\bfC}\ket{\psi_{2}}_{\bfS} \right).
\end{equation}
The expression for this output state demonstrates the core principle of the interferometer: the two paths of the interferometer (the $\ket{0}_{\bfC}$ and $\ket{1}_{\bfC}$ branches) contain different evolutions of $\bfS$.
Its inclusion in the circuit is crucial for creating the out-of-time-ordered structure of operators.

In order to extract $F_{\beta}(\tau)$ from output state $\ket{\Psi_{\text{out}}}$, we perform a measurement on $\bfC$.
By measuring the expectation values of the Pauli operators $\sigma^{x}$ and $\sigma^{y}$ on $\bfC$, we can directly determine the real and imaginary parts of $F_{\beta}(\tau)$, i.e.,
\eq{
F_{\beta}(\tau)=\expt{\sigma^{x}}_{\bfC}+\i\expt{\sigma^{y}}_{\bfC},
}
where
\eq{
\expt{\sigma^{x}}_{\bfC}&=\Re\left[\bra{\psi}V^{\dagger}(0)W(\tau)V(0)W^{\dagger}(\tau)\ket{\psi}\right]\nonumber\\
&=\Re F_{\beta}(\tau),\\
\expt{\sigma^{y}}_{\bfC}&=\Im\left[\bra{\psi}V^{\dagger}(0)W(\tau)V(0)W^{\dagger}(\tau)\ket{\psi}\right]\nonumber\\
&=\Im F_{\beta}(\tau).
}
Therefore, the OTOC can be calculated from Eq.~\eqref{eq:otoc-correlator} using $F_{\beta}(\tau)$.
The RTM was used in Ref.~\cite{Mi21} for measuring the time-dependent evolution and fluctuation of the OTOC to experimentally investigate the dynamics of quantum scrambling.
For the mixed state case as the input state, the specific generalization was proposed using the thermal field double state at finite temperature \cite{Sundar22}; and it was employed in Ref.~\cite{Green22}

\subsection{\label{subsec:wmm}Weak-measurement method}
\begin{figure*}[tb]
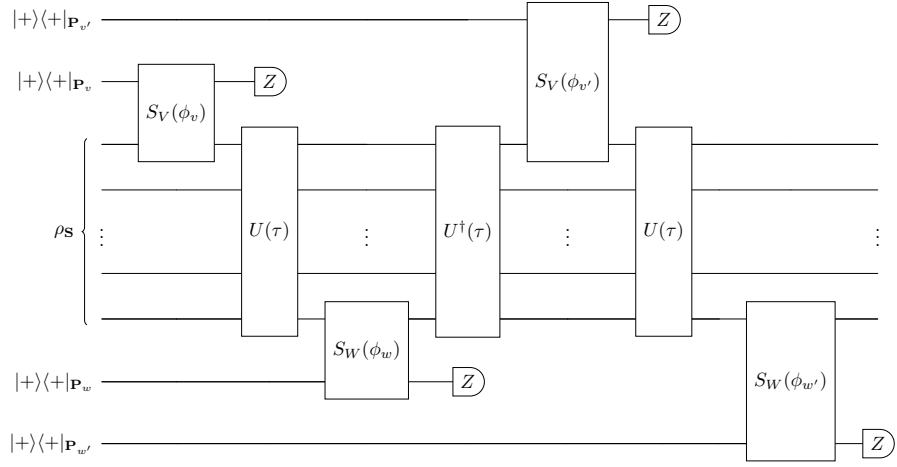

\centering
\usebox\boxB
\caption{
Schematic diagram of the weak-measurement method (WMM).
The WMM is a sequence of operations designed to evaluate the squared commutator $C_{\beta}(\tau)$.
The protocol consists of alternating weak measurements and time evolutions:
A weak measurement $M^{V}_{v}(\phi_{v})$ via the weak interaction $S_{V}(\phi_{v})=\exp\left[-\i\phi_{v}(V\otimes Y)/2\right]$, The forward time evolution $U(\tau)$, A second weak measurement $M^{W}_{w}(\phi_{w})$ using the weak interaction $S_{W}(\phi_{w})=\exp\left[-\i\phi_{w}(W\otimes Y)/2\right]$.
Apply the backward time evolution $U^{\dagger}(\tau)$, a third weak measurement $M^{V}_{v'}(\phi_{v'})$ via the weak interaction $S_{V}(\phi_{v'})=\exp\left[-\i\phi_{v'}(V\otimes Y)/2\right]$, the forward time evolution $U(\tau)$, and a fourth weak measurement $M^{W}_{w'}(\phi_{w'})$ using the weak interaction $S_{W}(\phi_{w'})=\exp\left[-\i\phi_{w'}(W\otimes Y)/2\right]$.
Finally, $C_{\beta}(\tau)$ is determined by averaging the products of the modified eigenvalues $\alpha_{v}(\phi_{v}),\alpha_{w}(\phi_{w}),\alpha_{v'}(\phi_{v'}),\alpha_{w'}(\phi_{w'})$ obtained from these measurements.
}\label{fig:wmm}
\end{figure*}

The WMM offers an alternative approach for measuring the OTOC that elegantly circumvents the need for backward time evolution \cite{Halpern17,Halpern18,Dressel18}.
This approach relates the OTOC to a sequence of weak measurements, where the information is extracted from the target system with minimal disturbance.
The method is particularly powerful when the operators of interest, $W$ and $V$, are unitary and satisfy the condition $W^{2} = \openone$ and $V^{2} = \openone$, which is called the dichotomic operator, such as Pauli operators, since it can be implemented with generalized measurements of any strength, not just weak ones.
This property eliminates the issue of disturbance typically associated with weak measurements \cite{González-Alonso19,Mohseninia19}.

The core idea is to measure the expectation values of nested commutators and anticommutators of the operators at different times.
These values can be extracted by performing a sequence of weak measurements and averaging the outcomes with specific weights.
To access the complex nature of the OTOC, we need a canonical type of measurement
\eq{
M^{A}_{\pm}(\phi)&:=\bra{z\pm}S_{A}(\phi)\ket{x+}\nonumber\\
&=\frac{1}{\sqrt{2}}\left[\cos\left(\frac{\phi}{2}\right)\openone\pm\sin\left(\frac{\phi}{2}\right)A\right],
}
where
\eq{
S_{A}(\phi)=\exp\left(-\i\frac{\phi}{2}A\otimes Y\right)
}
for an arbitrary observable $A$ and we use the notations $\ket{\bullet\pm}$ for $\bullet=x,y,z$ so that we can distinguish the bases of Pauli operators.
$M^{A}_{\pm}(\phi)$ causes a partial collapse of the state onto the eigenbasis of an operator $A$.
The outcome of the measurement $M^{A}_{\pm}(\phi)$ is weighted by modified eigenvalues that correspond to the anticommutator $\{A,B\}:=AB+BA$.
The real part of the correlator is extracted by performing a sequence of the measurements and averaging the results with specific weights.

For clarity, we now replace the $\pm$ notation with explicit labels, e.g., $\pm1\to(-1)^{a}$ with $a\in\{0,1\}$, which indicate the experimental outcome obtained when measuring the ancilla basis.
The strength of the measurement $M^{A}_{a}(\phi)$ is controlled by a coupling-strength $\phi\in(0,\pi/2]$ that ranges from a near-identity transformation to a projective measurement, allowing the tuning of the backaction to the system from weak $\phi=0$ to strong $\phi=\pi/2$.

We define the rescaled value that the experimenter should assign each observed ancilla outcome $a\in\{0,1\}$,
\eq{
\alpha_{a}(\phi)=\frac{(-1)^{a}}{\sin(\phi/2)}.
}
The values $\alpha_{a}(\phi)$ act as modified eigenvalues of the self-adjoint operator $A$, i.e., $A$ can be decomposed into the POVM for the measurement implemented by $M^{A}_{\pm}(\phi)$,
\eq{
\sum_{a=0,1}\alpha_{a}(\phi)[M^{A}_{a}(\phi)]^{\dagger}M^{A}_{a}(\phi)=A.
}
As a particularly important special case, when $\phi=\pi/2$, the values $\alpha_{a}(\phi)=(-1)^{a}$ reduce to the eigenvalues of $A$ and the measurements are projective ones.
These measurements can be implemented by coupling the system to an ancilla qubit with a controlled rotation gate and then measuring the ancilla. The strength of the measurement is controlled by a coupling angle $\theta \in (0, \pi/2]$. Weak measurements correspond to a small angle $\theta \to 0$.

Typically, determining complex quantities like operator correlators requires the use of weak measurements ($\phi\approx0$) to prevent disturbance.
In special cases, however, relevant information may still be contained in the collected measurement statistics in spite of any disturbance.
This is the case for qubits, where the following remarkable identities hold for
any coupling-strength and thus enable the improved correlator measurement protocols; the anticommutator identities
\eq{
\sum_{a=0,1}\alpha_{a}(\phi)M^{A}_{a}(\phi)\rho[M^{A}_{a}(\phi)]^{\dagger}&=\frac{1}{2}\{A,\rho\},\\
\sum_{a=0,1}\alpha_{a}(\phi)[M^{A}_{a}(\phi)]^{\dagger}BM^{A}_{a}(\phi)&=\frac{1}{2}\{B,A\}.
}
To measure $F_{\beta}(\tau)$, we consider the specific case where the operators $W$ and $V$ are dichotomic operators.
The protocol to obtain the real part of $F_{\beta}(\tau)$ is as follows.
\begin{enumerate}
\item Prepare each probe system $\bfP_{\bullet}$ with $\bullet=v,w,v',w'$ in the eigenstate $\ket{+}$ of the Pauli-$x$ operator $\sigma^{x}$ and $\bfS$ in the desired state $\rho_{\bfS}$.
\item Perform a weak measurement $M^{V}_{v}(\phi_{v})$ on the target system $\bfS$.
\item Evolve $\bfS$ forward in time by $U(\tau)$ for a duration from $t=0$ to $t=\tau$.
\item Perform a weak measurement $M^{W}_{w}(\phi_{w})$ on $\bfS$.
\item Evolve $\bfS$ backward in time by $U(\tau)$ for a duration from $t=0$ to $t=\tau$.
\item Perform a weak measurement $M^{V}_{v'}(\phi_{v'})$ on $\bfS$.
\item Evolve $\bfS$ forward in time by $U(\tau)$ for a duration from $t=0$ to $t=\tau$.
\item Perform a weak measurement $M^{W}_{w'}(\phi_{w'})$ on $\bfS$.
\end{enumerate}
Then, we average the products of the modified eigenvalues $\alpha_{v}(\phi_{v}),\alpha_{w}(\phi_{w}),\alpha_{v'}(\phi_{v'}),\alpha_{w'}(\phi_{w'})$ from each measurement.
Remarkably, the result of the averaging procedure directly yields a quantity related to the real part of $F(t)$ being of the form
\eq{
&\sum_{\substack{\alpha_{v}(\phi_{v}),\alpha_{w}(\phi_{w}),\\\alpha_{v'}(\phi_{v'}),\alpha_{w'}(\phi_{w'})\in\{0,1\}}}\alpha_{v}(\phi_{v})\alpha_{w}(\phi_{w})\alpha_{v'}(\phi_{v'})\alpha_{w'}(\phi_{w'})\nonumber\\
&\quad\mbox{}\times\Pr\{\alpha_{v}(\phi_{v}),\alpha_{w}(\phi_{w}),\alpha_{v'}(\phi_{v'}),\alpha_{w'}(\phi_{w'})\|\rho\}\nonumber\\
&\quad=\frac{1}{2^{3}}\expt{\{\{\{W(\tau),V(0)\},W(\tau)\},V(0)\}}_{\rho}\nonumber\\
&\quad=\frac{1}{2}(1+\Re[F_{\beta}(\tau)])\nonumber\\
&\quad=1-\frac{1}{4}C_{\beta}(\tau).
}
Note that this identity holds for any strength of the measurements.
Crucially, this method works even with strong, projective measurements, but the use of weak measurements allows for a reduced backaction on $\bfS$, which can be useful in certain experimental contexts.

\subsection{\label{subsec:ism}Irreversibility-susceptibility method}
\begin{figure}[tb]
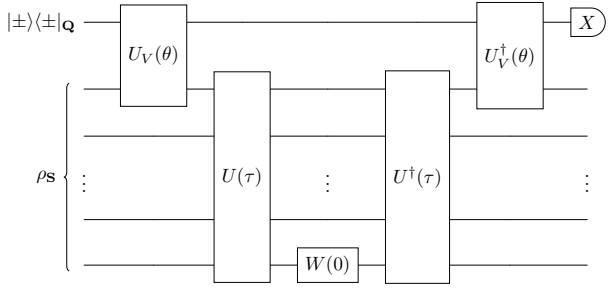

\centering
\usebox\boxC
\caption{
Schematic diagram of the irreversibility-susceptibility method (ISM).
The ISM is a protocol designed o evaluate the squared commutator $C_{\beta}(\tau)$ and consists of a sequence of operations:
A weak interaction $U_{V,\theta}=\exp\left[-\i\theta(Z\otimes V)\right]$, a scrambling process $\calD_{W}(\bullet):=W(\tau)(\bullet)W^{\dagger}(\tau)$, where $W(\tau)=U^{\dagger}(\tau)W(0)U(\tau)$ from $t=0$ to $t=\tau$, and a recovery map $\calR_{V,\bfS}:=\mathcal{J}_{\bfS}\circ\mathcal{U}^{\dagger}_{V,\theta}$, which is given by the inverse process of the weak interaction $\mathcal{U}_{V,\theta}$ and $\calJ_{\bfS}(\bullet):=\sum_{j=\pm}\bra{j}\Tr_{\bfS}(\bullet)\ket{j}\ketbra{j}{j}_{\bfQ}$.
Finally, the irreversibility of the entire process is calculated by comparing the initial and final states of the ancilla qubit system $\bfQ$.
}\label{fig:ism}
\end{figure}

The ISM leverages a recently developed formulation that connects the OTOC to the fundamental notion of irreversibility.
The core idea is that the OTOC, which quantifies quantum scrambling, can be reformulated as a specific measure of irreversibility of a quantum process [see the companion Letter \cite{Emori23} for more detail of the framework].
This formulation provides a thermodynamically meaningful perspective on the OTOC and offers an experimentally tractable way to evaluate it.

First, let us define the measure of irreversibility proposed in Ref.~\cite{Tajima22}, which is a useful tool to investigate universal limitations~\cite{Tajima22,Tajima25PRL,Nakajima24,Tajima25ARX} on the dynamics in various fields including black holes \cite{Hayden07,Yoshida19,Yoshida17,Yan20JUL,Tajima21,Schuster22,Brown23,Nakata23,Tajima22}, quantum thermodynamics \cite{Faist20,Kubica21,Tajima22,Tajima25PRL}, quantum measurements \cite{Wigner52,Araki60,Yanase61,Ozawa02M,Marvian12,Ahmadi13,Korzekwa13,Tajima19,Nakajima24,Tajima22}, quantum computing \cite{Ozawa02C,Karasawa09,Tajima18,Tajima20,Tajima22}, and channel costs of general resource theories \cite{Tajima25ARX}.
We consider a quantum process represented by a completely positive trace-preserving (CPTP) map $\calL$ that takes a state from a target system $\bfS$ to another system $\bfS'$.
For a given ensemble of input states $\Omega=\{p_{k},\rho_{k}\}$, where $\{\rho_{k}\}$ is a family of quantum states prepared with a probability distribution $\{p_{k}\}$, the irreversibility of the process $\calL$ is defined by
\eq{
\delta(\calL,\Omega)&:=\min_{\calR:\bfS'\ra \bfS}\sqrt{\sum_{k}p_{k}\delta^{2}_{k}},\label{eq:irreversibility}
}
as the minimum recovery error over all possible recovery maps $\calR$.
Here, the minimization is performed over all CPTP maps $\calR$, which attempt to reverse the process $\calL$.
$\delta_k$ quantifies the distance between the original state $\rho_k$ and the state after the loss and recovery processes $\calR\circ\calL(\rho_{k})$:
\eq{
\delta_{k}&:=D_{F}(\rho_{k},\calR\circ\calL(\rho_{k})),\label{eq:distance}
}
where $D_{F}(\rho,\sigma):=\sqrt{1-F(\rho,\sigma)^{2}}$ is the purified distance \cite{Tomamichel12}, which is based on the Uhlmann fidelity $F(\rho,\sigma):=\Tr\left(\sqrt{\sqrt{\rho}\sigma\sqrt{\rho}}\right)$ \cite{Uhlmann76}.
This definition of irreversibility has the benefit, providing operational meanings of a variety of irreversibility measures: it gives a minimum threshold for the entropy production \cite{Funo18} in stochastic thermodynamics, a lower bound for the entanglement fidelity error \cite{Watrous18} in quantum error correction, and a recovery error of the Petz recovery map \cite{Hayden04,Wilde15,Junge18}.

The OTOC is defined by the following expression, which directly relates it to a specific instance of irreversibility:
\eq{
C_{\beta}(\tau)&=\lim_{\theta\ra0}\frac{\delta^{2}(\mathcal{L}_{\rho,V,\theta,\calD_{W}},\Omega_{1/2,\pm},\calR_{V,\bfS})}{\theta^{2}},\label{eq:specific_dist=otoc}
}
Next, we break down the components of this formula.
The process $\calL_{\rho,V,\theta,\calD_{W}}$ is constructed from a composite map that simulates the time evolution under the dynamics of the OTOC:
\eq{
\calL_{\rho,V,\theta,\calD_{W}}:=(\calD_{W}\otimes\openone_{\mathbf{Q}})\circ\mathcal{U}_{V,\theta}\circ\mathcal{A}_{\rho},
}
which involves
\begin{enumerate}
\item Apply an appending process $\calA_{\rho}(\bullet):=\bullet\otimes\rho$ that prepares the initial state $\rho$ of $\bfS$ on an ancilla qubit system $\mathbf{Q}$, where the initial state of $\bfQ$ is prepared from a specific test ensemble $\Omega_{1/2,\pm}$.
\item Apply a weak interaction $\mathcal{U}_{V,\theta}(\bullet):=U_{V}(\theta)(\bullet)U^{\dagger}_{V}(\theta)$ on $\bfS+\bfQ$, where $U_{V}(\theta)=\exp\left[-\i\theta (Z\otimes V)\right]$.
\item Apply a scrambling process $\calD_{W}(\bullet):=W(\tau)(\bullet)W^{\dagger}(\tau)$, where $W(\tau)=U^{\dagger}(\tau)W(0)U(\tau)$.
\item Apply a recovery map $\calR_{V,\bfS}:=\mathcal{J}_{\bfS}\circ\mathcal{U}^{\dagger}_{V,\theta}$, which is given by the inverse process of the weak interaction $\mathcal{U}_{V,\theta}$ and $\calJ_{\bfS}(\bullet):=\sum_{j=\pm}\bra{j}\Tr_{\bfS}(\bullet)\ket{j}\ketbra{j}{j}_{\bfQ}$.
\end{enumerate}
Then, we calculate the irreversibility of the above process based on Eq.~\eqref{eq:irreversibility} and post-process its results using Eq.~\eqref{eq:specific_dist=otoc} to obtain $C_{\beta}(\tau)$.
Here, we assume that $W$ is a self-adjoint and unitary operator ($W=W^{\dagger}, WW^{\dagger}=\openone$) and $V$ is a self-adjoint operator.
Note that we can relax the condition of $W$ in general, while the condition is for the ease of the experimental implementation of the evaluation (see the companion Letter \cite{Emori23} and its Supplemental Material).
The term $\Omega_{1/2,\pm}$ prepares the initial states $\ket{\pm}$ with the probabilities $1/2$ in $\bfQ$.
In the numerical experiments, we only prepare the initial state $\ket{+}$ of $\bfQ$ and measure the Pauli-$x$ operator in the output state $\rho'$ of $\bfQ$ so that we can calculate the irreversibility
\eq{
C_{\beta}(\tau)=\lim_{\theta\ra0}\frac{\expt{\sigma^{x}}_{\rho}-\expt{\sigma^{x}}_{\rho'}}{2\theta^{2}},
}
as an equivalent and more experimentally convenient way \cite{Hofmann21,Emori24}.

Essentially, we are asking: how difficult is it to recover the initial state after it has been subjected to a weak perturbation $V(0)$ and a scrambling process $W(\tau)$?
The OTOC is shown to be proportional to the second-order term of the generated irreversibility.
The method offers a powerful framework for evaluating the OTOC on a quantum computer.
Unlike many traditional approaches that require complex interferometry or multiple-point measurements, the irreversibility-susceptibility method simplifies the experimental setup by requiring only a single final measurement on $\bfQ$.
This method makes it a highly promising and resource-efficient technique for practical implementations and can be implemented by measuring the distinguishability of quantum states, which is a natural task for quantum information processing.

\begin{figure}[tb]
\centering
\includegraphics[width=\linewidth]{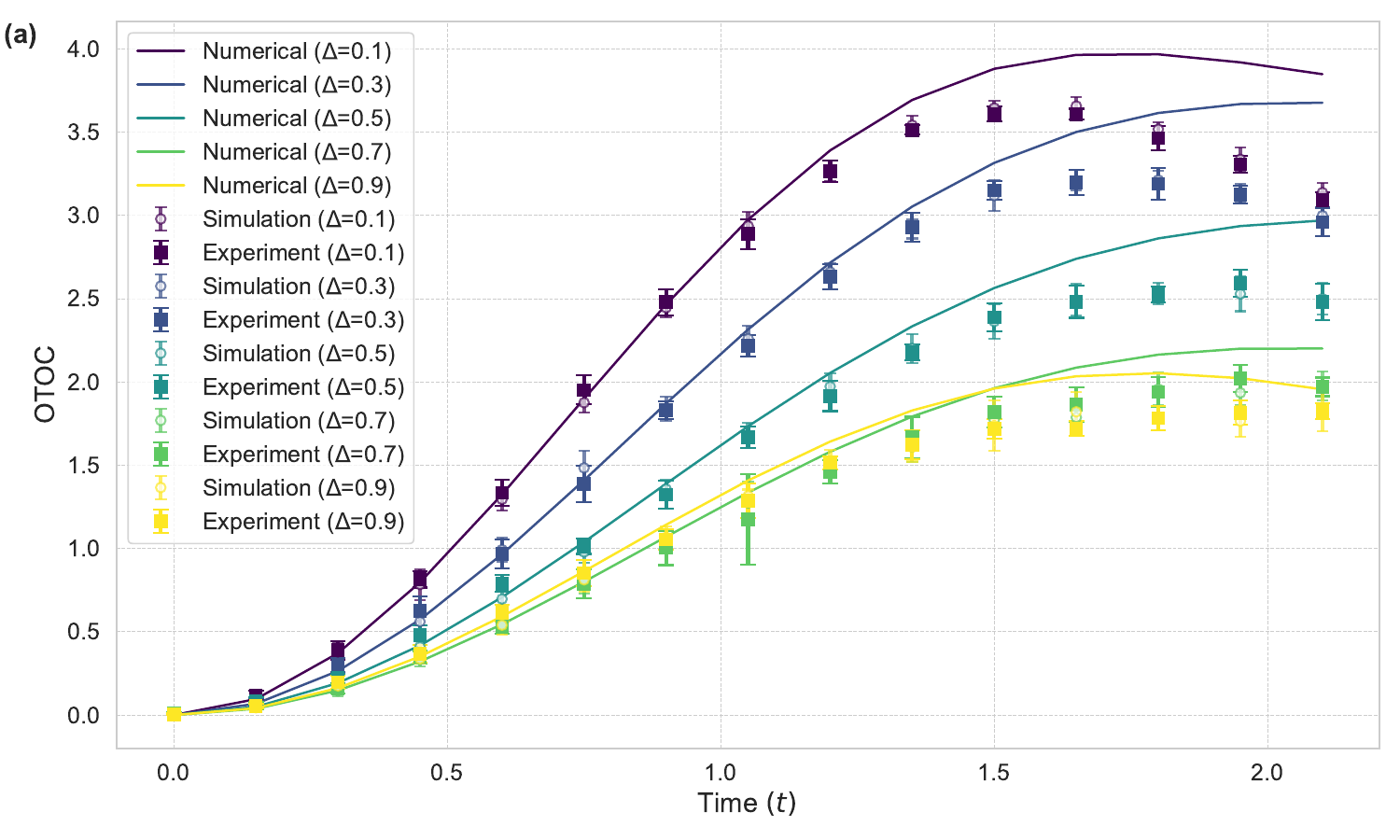}
\includegraphics[width=\linewidth]{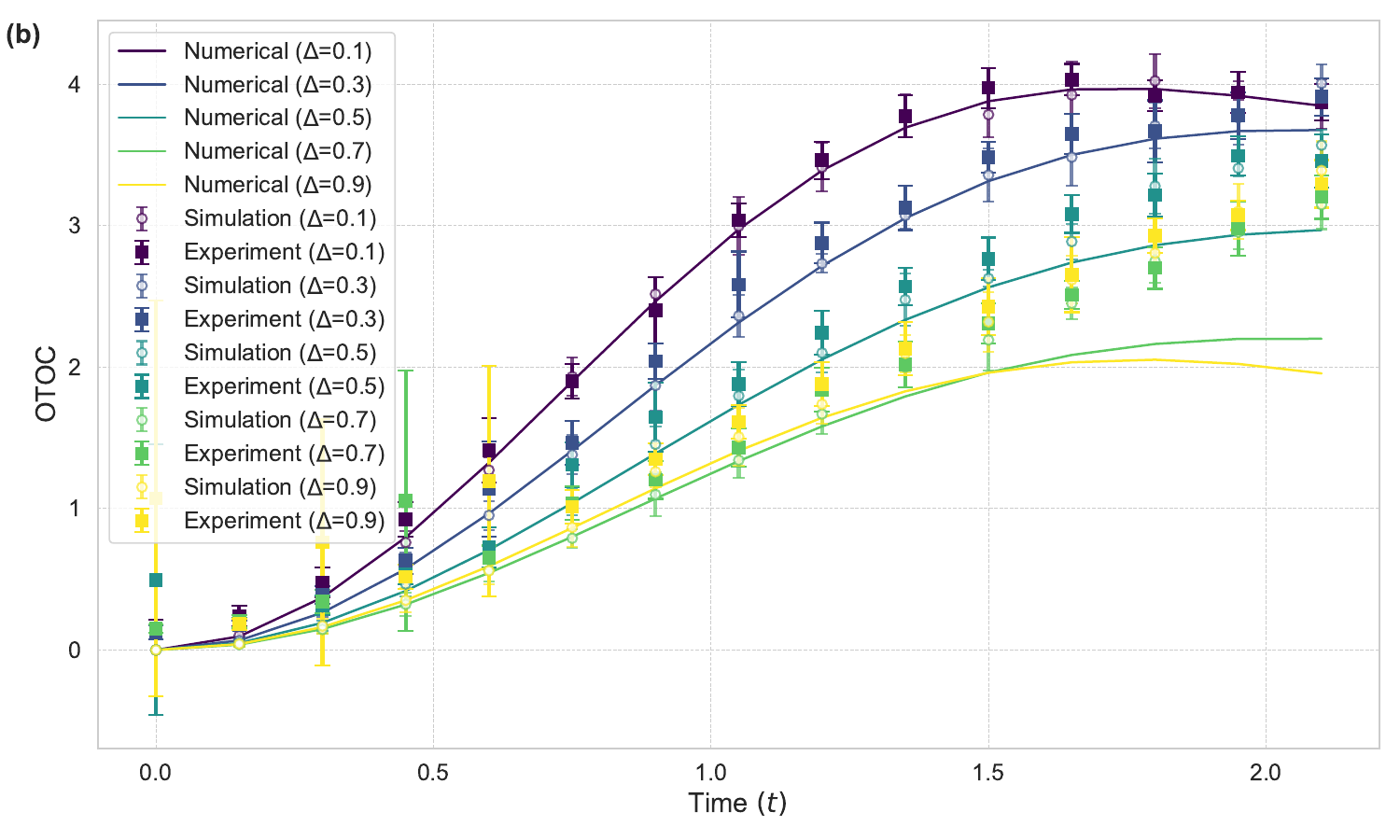}
\includegraphics[width=\linewidth]{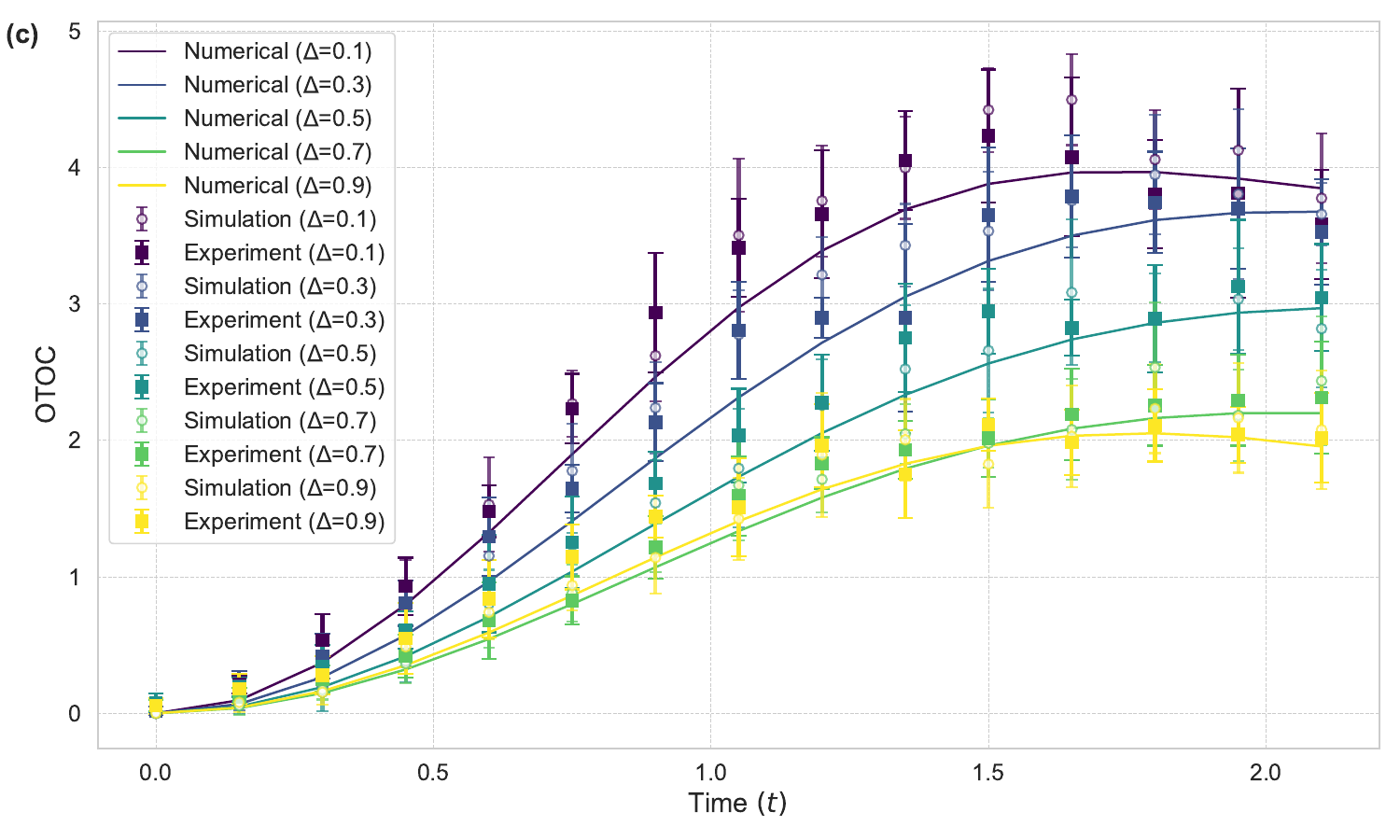}
\caption{\label{fig:res_otoc}
The experimental results of the measurement of the OTOC.
The results present a comparison among the ideal values (solid lines) from matrix calculations, the noiseless values (circle dots) from \texttt{aer-simulator}, and the experimental results (square dots) from \texttt{reimei}.
The error bars represent the standard deviation of the measured values.
While the experimental data show good overall agreement with the theoretical predictions, clear method-dependent behaviors of the OTOC are observed.
(a) is the OTOC evaluated using the rewinding time method (RTM).
(b) is the OTOC evaluated using the weak-measurement method (WMM).
(c) is the OTOC evaluated using the irreversibility-susceptibility method (ISM), which shows a larger standard deviation because of its reliance on the weak interactions.
}
\end{figure}

\section{\label{sec:ne}Experiments on the trapped-ion quantum computer, \texttt{reimei}}
To overcome the challenges of measuring the OTOC, many experiments have been conducted with various ingenious techniques \cite{Li17,Gärttner17,Aggarwal18,Landsman19,Khurana19,Meier19,Sieberer19,Niknam20,Chen20,Nie20,Joshi20,Braumüller22,Geller22,Harris22,Green22,Zhao22,Wang22,Kuper22,Xiang24,Liang25,Huang25,Google25}.
However, these experiments have primarily focused on analyzing the properties of the OTOC within different models, with little to no consideration given to how the experimental evaluation methods themselves influence the results.
Therefore, comparing the performance of the various methods used to measure the OTOC is crucial for a deeper understanding of its intrinsic properties.
In this study, we compare the RTM, WMM, and ISM to investigate the method-dependent properties of the OTOC.
We conducted all experiments on the trapped-ion quantum computer---Quantinuum \texttt{reimei} \cite{Quantinuum}, which is accessible through the Quantinuum Nexus.
We note that this work represents the first experimental application of both the WMM and ISM, using the actual hardware.

\subsection{\label{subsec:setup}Setup}
We performed numerical experiments on the evaluation of the OTOC using a cloud-based quantum computer; we conducted all experiments on the trapped-ion quantum computer---Quantinuum \texttt{reimei} \cite{Quantinuum}.
The setup for our experiment, including the preparation of the initial state, is detailed below.

\subsubsection{\label{subsubsec:ini_state}Initial state}
For the initial state, we chose to use a Gibbs state.
For a given Hamiltonian $H$ and an inverse temperature $\beta = 1/(k_{B}T)$, the Gibbs state is defined as
\eq{
\rho(\beta,H)=\frac{\e^{-\beta H}}{Z(\beta,H)}.\label{eq:gibbs}
}

Preparing such a state on a quantum computer is a non-trivial task.
This is because the state is generally a mixed state, and preparing a mixed state is not possible with only unitary operations on a closed system.
To overcome this challenge, we utilized a variational quantum algorithm (VQA) proposed in Ref.~\cite{Consiglio24}.
This algorithm prepares a purified version of the Gibbs state on an enlarged system consisting of both a target system $\bfS$ and an ancilla system $\bfA$.
The key technique of this algorithm is based on the principle that the Gibbs state is the unique state that minimizes the Helmholtz free energy for a given Hamiltonian.
The free energy is of the form
\eq{
F(\rho)=\Tr(H\rho)-\beta^{-1}S(\rho),
}
where
\eq{
S(\rho):=-\Tr(\rho\log\rho)
}
is the von Neumann entropy.
The Gibbs state is the solution to the minimization problem
\eq{
\rho_{\text{Gibbs}}=\arg\min_{\rho}F(\rho).
}
A significant challenge in realizing the minimization on a quantum computer is the direct measurement of the von Neumann entropy $S(\rho)$, which is not a simple observable.
The VQA we employed cleverly avoids this issue by preparing a special kind of entangled state, called a thermofield double (TFD) state, on $\bfA+\bfS$.
The algorithm uses a parameterized quantum circuit (PQC) composed of two parameterized unitaries $U_{\bfA}(\bm{\theta})$ and $U_{\bfS}(\bm{\phi})$.
The unitary $U_{\bfA}$ acts on the ancilla qubits $\bfA$, and $U_{\bfS}$ acts on the target qubits $\bfS$.
These two parts are connected by a series of CNOT gates between corresponding qubits of the two registers.

Starting from the initial state $\ket{0}^{\otimes 2n}$, where $n$ is the number of qubits in each register, the circuit performs the following operations.
\begin{enumerate}
\item Apply the unitary operator $U_{\bfA}(\bm{\theta})$ to $\bfA$:
\eq{
[U_{\bfA}(\bm{\theta})\otimes\openone_{\bfS}]\ket{0}^{\otimes 2n}=\ket{\Psi(\bm{\theta})}_{\bfA}\otimes\ket{0}^{\otimes n}_{\bfS},
}
where $\ket{\Psi(\bm{\theta})}_{\bfA}=\sum_{i=0}^{2^{n-1}}\sqrt{p_{i}(\bm{\theta})}\ket{\psi_{i}(\bm{\theta})}_{\bfA}$ and we denote by $\{\ket{\psi_{i}(\bm{\theta})}\}$ by the computational basis on a quantum computer.
The coefficients $\sqrt{p_{i}(\bm{\theta})}$ are determined by the parameters $\bm{\theta}$ of the unitary operator $U_{\bfA}(\bm{\theta})$.
\item Apply CNOT gates between $\bfA+\bfS$.
The CNOT gates entangle the two registers, creating the state
\eq{
\ket{\text{TFD}(\bm{\theta})}=\sum_{i=0}^{2^{n-1}}\sqrt{p_{i}(\bm{\theta})}\ket{\psi_{i}(\bm{\theta})}_{\bfA}\ket{\psi_{i}(\bm{\theta})}_{\bfS}.
}
This state is the TFD state in the computational basis.
\item Apply the unitary operator $U_{\bfS}(\bm{\phi})$ to $\bfS$:
\eq{
\ket{\widetilde{\text{TFD}}(\bm{\theta},\bm{\phi})}&=[\openone_{\bfA}\otimes U_{\bfS}(\bm{\phi})]\ket{\text{TFD}(\bm{\theta})}\nonumber\\
&=\sum_{i=0}^{2^{n-1}}\sqrt{p_{i}(\bm{\theta})}\ket{\psi_{i}(\bm{\theta})}_{\bfA}\ket{E_{i}(\bm{\theta},\bm{\phi})}_{\bfS}.
}
Here, we assume that $U_{\bfS}(\bm{\phi})$ transforms the computational basis states $\ket{\psi_{i}(\bm{\theta})}$ to the eigenstates $\ket{E_{i}(\bm{\theta},\bm{\phi})}$ of $H$.
\end{enumerate}
By tracing out the register $\bfA$ from the final TFD state, we obtain a mixed state on $\bfS$:
\eq{
\rho_{\bfS}(\bm{\theta},\bm{\phi})&=\Tr_{\bfA}[\ketbra{\widetilde{\text{TFD}}(\bm{\theta},\bm{\phi})}{\widetilde{\text{TFD}}(\bm{\theta},\bm{\phi})}]\nonumber\\
&=\sum_{i=0}^{2^{n-1}}p_{i}(\bm{\theta})\ketbra{E_{i}(\bm{\theta},\bm{\phi})}{E_{i}(\bm{\theta},\bm{\phi})}_{\bfS},
}
where $p_{i}(\bm{\theta})=$
The crucial insight of this algorithm is that we can get the probability distribution $p_{i}(\bm{\theta})$ to calculate $S(\rho)$ by measuring $bfA$ in the computational basis without directly measuring $\bfS$, i.e., applying the tracing out operation over $\bfS$ to  $\ket{\text{TFD}(\bm{\theta})}$ yields a mixed state of $\bfA$:
\eq{
\rho_{\bfA}(\bm{\theta})&=\Tr_{\bfS}[\ketbra{\text{TFD}(\bm{\theta})}{\text{TFD}(\bm{\theta})}]\nonumber\\
&=\sum_{i=0}^{2^{n-1}}p_{i}(\bm{\theta})\ketbra{\psi_{i}(\bm{\theta})}{\psi_{i}(\bm{\theta})}_{\bfA}.
}
This property means that the von Neumann entropy $S(\rho_{\bfS})$ of $\bfS$ is equal to the von Neumann entropy $S(\rho_{\bfA})$ of $\bfA$.
Since $\rho_{\bfA}$ is a diagonal state in the computational basis, its entropy can be calculated simply from $\{p_{i}(\bm{\theta})\}$.
Thus, we can avoid the need for complex task---direct entropy measurements on the quantum computer.

The cost function of the VQA is therefore defined as the free energy of $\bfS$, which is minimized by optimizing the parameters $\bm{\theta}$ and $\bm{\phi}$:
\eq{
F(\bm{\theta},\bm{\phi})=\Tr[H\rho_{\bfS}(\bm{\theta},\bm{\phi})]-\beta^{-1}S(\rho_{\bfA}(\bm{\theta})).
}
The first term can be measured on $\bfS$.
The second term can be calculated classically from $\{p_{i}(\bm{\theta})\}$.
The VQA iteratively optimizes the parameters to minimize $F(\bm{\theta},\bm{\phi})$ until convergence.
Upon finding the optimal parameters $\bm{\theta}^{*}$ and $\bm{\phi}^{*}$, the circuit prepares the Gibbs state $\rho_{\text{Gibbs}}\approx\rho_{\bfS}(\bm{\theta}^{*},\bm{\phi}^{*})$ on $\bfS$.

\subsubsection{\label{subsubsec:gen}Observables}
In our experiment, we chose the one-dimensional Heisenberg XXZ model in a transverse magnetic field as the generator of dynamics for our system of interest.
This is a canonical model in condensed matter physics known for its rich phase diagram, which includes both integrable and chaotic regimes.
The Hamiltonian is given by
\eq{
H=-\frac{1}{4}\sum^{n-1}_{i=1}(\sigma^{x}_{i}\sigma^{x}_{i+1}+\sigma^{y}_{i}\sigma^{y}_{i+1}+\Delta\sigma^{z}_{i}\sigma^{z}_{i+1})-h\sum^{n}_{i=1}\sigma^{z}_{i},\label{eq:xxz}
}
where $\sigma^{\bullet}_{i}$ for $\bullet=x,y,z$ are the Pauli matrices acting on the $i$-th qubit, $\Delta$ is the anisotropy parameter, and $h$ is the strength of the transverse magnetic field.
For the present paper, we set the magnetic field strength to $h=(1-\Delta)/2$.

Our numerical experiments were conducted with the following specific parameters:
\begin{description}
\item[The size of $\bfS$] $\bfS$ consisted of $n=4$ qubits.
\item[The parameters of $H$] We vary the anisotropy parameter $\Delta$ from $0.1$ to $0.9$: $\Delta=\{0,1,0.3,0.5,0.7,0.9\}$.
\item[The time evolution] We measured the evolution of the OTOC from time $t=0$ up to $t=2.1$.
This time interval was discretized into $15$ time steps.
\item[Observables] For the evaluation, we selected local Pauli operators as our observables.
Specifically, we chose both $W(0)$ and $V(0)$ to be the Pauli-$x$ operator on the first qubit: $W(0)=V(0)=\sigma^{x}_{1}$.
In terms of the full system, this corresponds to the operator $X\otimes\openone\otimes\openone\otimes\openone$.
\end{description}
The value of the measured OTOC at each time step was averaged over $10$ independent experimental iterations to mitigate statistical noise.
For the statistical averaging, each quantum circuit was executed with a total of $1000$ measurement shots.
In the ISM, the interaction strength was set to $\theta=0.4$ radians.
The initial state for our experiment was prepared as an approximation of the Gibbs state defined in Eq.~\eqref{eq:gibbs}, with the XXZ Hamiltonian specified in Eq.~\eqref{eq:xxz}.

\subsection{\label{subsec:res}Results}
As shown in Fig.~\ref{fig:res_otoc}, the solid lines represent the ideal values obtained from numerical calculation.
The circle dots show the values from noiseless simulations via \texttt{aer-simulator}, while the square dots represent the experimental data obtained from the \texttt{reimei}.
The error bars indicate the standard deviation of the measured values.

Overall, a consistent observation across all the results obtained from three methods is the subtle deviation between the simulation and experimental results, which is primarily attributed to errors from extrinsic noise.
In addition, the discrepancies between the theoretical and the simulation results are likely due to a combination of two factors: fidelity errors from potential imperfections in the preparation of the variational Gibbs state, and algorithmic errors from finite-value truncation in the Trotter decomposition.
Despite these inaccuracies, the results clearly demonstrate the feasibility of experimentally exploring quantum dynamics on current quantum computers.

Next, we compare the results for each method in detail.
A notable difference among the methods is the variation in the behavior of the measured OTOC values with respect to the parameter $\Delta$ in the XXZ Hamiltonian.
Figure \ref{fig:res_otoc}(a) shows the OTOC evaluated using the RTM.
At early times, the results align well with the ideal values regardless of the values of $\Delta$.
However, as time progresses, the measured values for $\Delta = 0.5$, $0.7$, and $0.9$ deviate downward from the theoretical predictions.
Figure \ref{fig:res_otoc}(b) shows the OTOC evaluated using the WMM.
Similar to the RTM, the results match the ideal values at early times.
In the later time regime, however, the values for $\Delta = 0.5$, $0.7$, and $0.9$ deviate upward from the theoretical predictions.
Figure \ref{fig:res_otoc}(c) shows the OTOC evaluated using the ISM.
For all values of $\Delta$, the measured OTOC is in good agreement with the theoretical values on average.
Because of the use of a weak interaction, the standard deviation for the ISM is larger compared to the other two methods.
However, we must carefully interpret the apparent robustness of the ISM against variations in the anisotropy parameter $\Delta$.
As observed in Fig.~\ref{fig:res_otoc}(c), the ISM exhibits larger standard deviations compared to the RTM and WMM.
This variance is not a hardware artifact, but an intrinsic feature of protocols relying on weak interactions.
Specifically, the useful signal extracted in the ISM---the shift in the expectation value of the ancilla's Pauli-$x$ operator---scales quadratically with the weak interaction strength: $1 - \expt{\sigma^{x}}_{\rho'} \approx 2\theta^{2}C_{\beta}(\tau)$.
Conversely, the shot noise for $N$ measurement shots scales as $1/\sqrt{N}$. Consequently, the signal-to-noise ratio (SNR) is inherently suppressed by the weak coupling parameter, yielding $\text{SNR} \sim \mathcal{O}(\theta^{2}\sqrt{N})$. 
In our experiment, with $\theta = 0.4$ radians and $N = 1000$ shots, the statistical noise floor remains relatively high.
Therefore, it is plausible that any subtle, method-dependent deviations induced by the physical parameter $\Delta$---which are clearly resolved in the $\mathcal{O}(1)$ signals of the RTM and WMM---are currently being masked by the dominant statistical uncertainty in the ISM.
While the mean values of the ISM faithfully track the ideal values across all tested regimes, this fundamental variance penalty precludes a definitive conclusion regarding the absence of $\Delta$ dependence.
Overcoming this limitation in future experiments will require significantly increasing the shot count $N$ to compensate for the $\theta^{2}$ suppression.

In summary, while the ISM shows little dependence on the parameter $\Delta$, both the RTM and WMM exhibit reduced accuracy in the later time regime for larger values of $\Delta$.
This method-dependent behavior of the measured OTOC properties has not been previously reported.
Further investigation and detailed analysis are needed to understand the underlying causes of this phenomenon.

\section{\label{sec:conc}Conclusion}
In this study, we have presented the numerical evaluations of the OTOC on the \texttt{reimei} using three distinct protocols: the rewinding time method (RTM), the weak-measurement method (WMM), and the irreversibility-susceptibility method (ISM).
By successfully implementing these methods on the \texttt{reimei} with a variationally prepared Gibbs state of the XXZ Heisenberg model, we have demonstrated the practical feasibility of exploring complex quantum dynamics on current-generation hardware.
Our experimental results show a good overall agreement with theoretical predictions, validating these methods as practical tools for investigating quantum chaos.

A central aspect of our work was the comparative analysis of these three experimental schemes.
We found that each method has unique strengths and limitations.
The RTM is conceptually straightforward but requires precise measurement of two observables.
The WMM offers physical insight into the OTOC based on quasiprobability, though it can be demanding in terms of the number of weak measurements required.
The ISM provides a direct link between scrambling and irreversibility, but its performance is highly sensitive to the gate fidelity of the weak interactions.

The significance of our findings is twofold.
First, we have provided the first experimental demonstration of the ISM for measuring the OTOC in a non-trivial many-body system, thereby expanding the toolkit of protocols for studying quantum information scrambling.
Second, our work contributes to the growing evidence that current quantum processors are effective platforms for exploring complex, fundamental phenomena in quantum information science.

Our results also highlight a previously unreported, method-dependent behavior of the measured OTOC with respect to the XXZ Hamiltonian parameter $\Delta$, particularly for the RTM and WMM.
Future work should focus on clarifying the underlying cause of this dependence.
Furthermore, to probe the onset of many-body chaos more directly, it is essential to scale these experiments to a larger number of qubits and investigate how the OTOC depends on system size.
Applying these methods to different Hamiltonians would also provide deeper insights into the nature of quantum dynamics.
Finally, improving the fidelity of variational Gibbs state preparation and incorporating advanced error mitigation techniques will be crucial steps toward achieving quantitatively precise OTOC measurements on near-term quantum hardware.

\begin{acknowledgments}
HE was supported by JST SPRING, Grant Number JPMJSP2119, and RIKEN Junior Research Associate Program.
HT was supported by JSPS Grants-in-Aid for Scientific Research 
No. JP25K00924, and MEXT KAKENHI Grant-in-Aid for Transformative
Research Areas B ``Quantum Energy Innovation'' Grant Numbers 24H00830 and 24H00831, JST MOONSHOT No. JPMJMS2061, and JST FOREST No. JPMJFR2365.
\end{acknowledgments}

\bibliography{ref}

\end{document}